\documentclass[pra,showpacs]{revtex4}
\newcommand{\rb}{\mbox{\boldmath $r$}}
\newcommand{\pb}{\mbox{\boldmath $p$}}

\newcommand{\Ob}{\mbox{\boldmath $0$}}

\newcommand{\be}{\begin{equation}}
\newcommand{\ee}{\end{equation}}

\begin{document}

\title{Relation between positional specific heat and static relaxation length: 
Application to supercooled liquids}

\author{S.\ Davatolhagh}
\affiliation{Department of Physics, College of Sciences,
 Shiraz University, Shiraz 71454, Iran}
\date{\today}

\begin{abstract}
A general identification of the {\em positional specific heat} as
the thermodynamic response function associated with the {\em
static relaxation length} is proposed, and a phenomenological
description for the thermal dependence of the static relaxation
length in supercooled liquids is presented. Accordingly,  through
a phenomenological determination of positional specific heat of
supercooled liquids, we arrive at the thermal variation of the
static relaxation length $\xi$, which is found to vary in
accordance with $\xi \sim (T-T_0)^{-\nu}$ in the quasi-equilibrium
supercooled temperature regime,
 where $T_0$ is the Vogel-Fulcher temperature and
exponent $\nu$ equals unity.  This result to a certain degree
agrees with that obtained from mean field theory of
random-first-order transition, which suggests a power law
temperature variation for $\xi$ with an apparent divergence at
$T_0$. However, the phenomenological exponent $\nu = 1$, is higher
than the corresponding mean field estimate (becoming exact in
infinite dimensions), and in perfect agreement with the relaxation
length exponent as obtained from the numerical simulations of the
same models of structural glass in three spatial dimensions.
\end{abstract} 

\maketitle


\section{INTRODUCTION}
\label{sec:1} The deepest and most interesting unsolved problem in
the theory of solids is probably that of the theory and the nature of
glass and glass transition \cite{And}. One of the most striking
features of the typical supercooled liquid is that its relaxation
time or viscosity $\eta$ changes by several decades on changing
the temperature by a few tens of degrees. All available data for
viscosity fall between Arrhenius and highly non-Arrhenius
extremes, designated `strong' and `fragile', respectively
\cite{Ang91}. The latter is characterized by a highly
temperature-dependent effective energy barrier against the viscous
flow; i.e., a temperature-dependent energy barrier $E_{\rm eff}$
appears in $\eta = \eta_0\exp (\beta E_{\rm eff})$, where $\eta_0$
is a temperature-independent but species-dependent parameter of
the orders of $10^{-2}$-$10^3$ poise, $\beta =1/k_B T$, and $k_B$
is the Boltzmann constant. The temperature variation of the
viscosity for fragile supercooled liquids is described accurately
over a wide range of temperatures by the Vogel-Fulcher empirical
equation \cite{VF}: \be \eta = \eta_0 \exp \left( \frac{A}{T-T_0}
\right). \label{eq:5.25} \ee The apparent divergence temperature
$T_0$ appearing in Eq.\ (\ref{eq:5.25}) is called the
Vogel-Fulcher temperature and is often found to be very close to
the Kauzmann temperature $T_K$ \cite{Kauz}, where a
configurational entropy of the liquid extrapolates to zero
\cite{Ang97}. This well known experimental fact, $T_0 \approx T_K$
is indeed a matter of considerable interest as it seems to suggest
that the ideal glass transition temperature observed dynamically,
and thermodynamically, must have common physical origin
\cite{Elliott}. It should be also pointed out that in addition to
non-Arrhenius variation of the viscosity with the temperature in
the supercooled temperature regime, fragile liquids are also
characterized by a distinct jump in the second-order thermodynamic
functions such as the specific heat $C_p$, the isothermal
compressibility $\kappa_T$, and the thermal expansion coefficient
$\alpha$ at the laboratory or calorimetric glass temperature
$T_g$, where $T_g > T_0$.

Furthermore, fragile supercooled liquids are also distinguished by
a highly non-exponential relaxation response as they approach
equilibrium when perturbed. Often the Kohlrausch-Williams-Watts
(KWW) \cite{KWW}, or stretched exponential function is used to
characterize the relaxation response of fragile liquids $\phi (t)
= \exp [-(t/\tau )^{\beta} ]$, where $\beta$ is the
non-exponentiality parameter such that $0<\beta <1$, $\tau$ is a
relaxation time, and both are found to be temperature-dependent.
The above relaxation response typical of fragile liquids is well
explained in terms of the existence of dynamically heterogeneous
regions in the supercooled liquid such that the relaxation in a
given region is exponential but the average relaxation time $\tau$
varies with a broad distribution among regions \cite{XW01}. This
dynamical heterogeneity of the supercooled liquids has been
further confirmed by the recent numerical \cite{Get00}, and
experimental research \cite{Ret01}, which confirm the presence of
such regions with characteristic lengths spanning 100s of
molecules. The above numerical and experimental confirmations of
the dynamical heterogeneity of the supercooled glass forming
liquids, lends further support to the notion of cooperatively
rearranging regions (CRRs) in a supercooled liquid \cite{Stan}.

There are several indications that the viscosity, and the various
other structural relaxation times of a supercooled liquid must be
correlated with the average size of a CRR, which is a concept
dating back to the considerations of cooperative relaxation by the
Adam and Gibbs \cite{AG}. In their approach the increase of the
effective potential energy barrier $E_{\rm eff} = z \Delta\mu$,
with $\Delta\mu$ being largely the potential energy barrier
against rearranging a single molecule in a cluster found by $z$
molecules, is due to an increase in the cluster size $z$ as the
temperature is lowered. This cooperativity concept requires a
characteristic static length $\xi$ characterizing the average
linear size of a CRR. Evidently, an incremental increase in $\xi$
of the order of a few nanometers is magnified in an exponentially
large (macroscopic) relaxation time as a consequence of which the
supercooled liquid falls out of equilibrium on experimental time
scales, hence, making any underlying static thermodynamic
transition unreachable under the laboratory conditions. From a
theoretical point of view, the mean field theory of
random-first-order transition (or discontinuous spin glasses) that
exhibits qualitative features in tandem with the structural glass
phenomenology, suggests an approximate power law temperature
dependence for the static relaxation length such that $\xi \sim
(T-T_0)^{-\nu}$, where the apparent divergence temperature $T_0$
is the Vogel-Fulcher temperature, and the exponent $\nu = 2/d$ or
$\nu = 2/3$ in $d=3$ dimensions \cite{Kirk,Donth91}. There has
been an empirical attempt to investigate the temperature variation
of $\xi$ the results of which are more or less consistent with the
above proposed power-law \cite{Fischer}. Results reported for
fragile liquid o-terphenyl are $\nu = 0.69\pm 0.06$ with $T_0 =
203 \pm 6$\ K. But the relevance to the static relaxation length
of the experimental procedure adopted in Ref.\ \cite{Fischer}, and
the various interpretations of the experimental data have been called
into question by the subsequent experimental investigations
\cite{Nagel}. More recently, numerical simulations in three
dimensions of the microscopic models that exhibit
random-first-order transition in the mean field such as, the
$p$-spin glasses \cite{CCP}, and the frustrated Ising lattice gas
model \cite{CC}, are found to be more in favour of a static relaxation
length exponent $\nu = 1$.

In an attempt to clarify some of the discrepancies concerning the
precise nature of the thermal dependence of the static relaxation
length $\xi$ of the fragile supercooled liquids, as alluded to in
the above discussion, we adopt a phenomenological approach to
obtain the temperature variation of $\xi$. The main ingredients in
this semi-empirical approach are: (i) the temperature-dependent
potential energy barrier $E_{\rm eff}$ against the viscous flow
that is embodied in the Vogel-Fulcher equation for the viscosity.
(ii) The thermodynamic response function bond susceptibility
$\chi_b$ as applied here to the case of liquids, which is to be
regarded as the response function measure of tendency for bond
ordering or correlated relaxation of bonds into their low-lying
energy states, brought about by the rearrangement of a CRR. Bond
susceptibility was introduced earlier in the context of a
diluted-bond model system relevant to the problem of glass
transition, the thermodynamic properties of which were
investigated by means of the Monte Carlo simulation \cite{D01}.
Here, the very concept underlying bond susceptibility, i.e.,
correlated ordering or relaxation of bonds where intermolecular
bonds are treated as distinct objects possessing internal degrees
of freedom or energy states, is generalized and applied to the
case of laboratory liquids. This approach paves the way for
identification of the interaction or positional specific heat
$C_i$ as the thermodynamic response function associated with the
characteristic length of relaxation $\xi$. Subsequently, a
semi-empirical determination of the positional specific heat $C_i$ for
the general class of fragile liquids, is used to arrive at the
thermal variation of static relaxation length $\xi$ that, by
definition, gives the average linear size of a CRR in the liquid.

The rest of this paper is organized as follows. In Sec.\
\ref{sec:2} we describe the relevant conceptual and theoretical
background concerning the various competing ordering processes in
a liquid, and give a definition for the bond susceptibility
$\chi_b$ as a response function measure of tendency for correlated
bond ordering. Sec.\ \ref{sec:3} contains the derivations of the
various relationships among thermodynamic and correlation
functions relevant to the present discussion. As it will become
evident in Sec.\ \ref{sec:3.1}, there exists a relationship of the
form $\chi_b = T \cdot C_i$ relating the bond susceptibility of a
liquid in canonical ensemble with the interaction or positional
part of the specific
heat. Furthermore, in Sec.\ \ref{sec:3.2} bond susceptibility is
shown to be intimately related to static relaxation length such that
essentially $\chi_b \sim \beta\xi^2$. These results essentially
point at association of positional specific heat $C_i$ as a
thermodynamic response function with characteristic length of
relaxation $\xi$, which is a novel concept brought to light in
section \ref{sec:3.3}. This association of $C_i$ and $\xi$
is then applied in
Sec.\ \ref{sec:4} to develop a phenomenological description for
the temperature variation of the static relaxation length in
fragile supercooled liquids. Concluding remarks and a summary of
the main results are presented in Sec.\ \ref{sec:5}.

\section{RELEVANT BACKGROUND}
\label{sec:2} \subsection{Two-order-parameter description of
liquids}\label{sec:2.1} Attempts have been made to incorporate
frustration arising from the local ordering of bonds in a
supercooled liquid through the introduction of a local order
parameter characterizing the energetically favoured local
arrangements of the liquid molecules, which are not consistent
with the crystallographic symmetry favoured by the density
ordering or crystallization. In this two-order-parameter
description of the liquids due to H.\ Tanaka \cite{T98,T99}, the
frustration arises from competition between density ordering and
local bond ordering, explaining why some molecules crystallize
easily without vitrification, while others easily form glasses
without crystallization. The effect of density ordering is to
maximize the density of molecules favouring a close-packed
crystallographic symmetry, while local bond ordering tends to
improve the quality of bonds by reducing the bond energies at the
local level. This model therefore emphasizes that introduction is
necessary of a bond order parameter, in addition to the density
$\rho(\rb)$, in order to have a complete thermodynamic description
of the liquid state, and, in particular, of the supercooled glass
forming liquids. The energetically favoured local structures, such
as, e.g., the icosahedral arrangements favoured by the spherical
molecules \cite{Frank}, are taken to be randomly distributed in a
sea of normal liquid. It is further argued that the local
structures with finite, but long life times, act as impurities and
produce the effects of fluctuating interactions and {\em
symmetry-breaking random fields} against density ordering in a
liquid, in much the same way as magnetic impurities frustrate
magnetic ordering in a spin glass system \cite{MP}. In this
two-order-parameter description of the liquids the `bond order
parameter' $ S(\rb)$ is taken to be defined by the local
concentration of the energetically favoured structures, and the
average concentration of local structures $\bar{S}$ is estimated
to be given by $\bar{S} \sim g_S / g_{\rho} \exp [\beta(E_{\rho} -
E_S)]$, where $E_i$ and $g_i$ are the energy level and the number
of degenerate states of the $i$-type structure. ($i$=$\rho$
corresponds to the normal liquid while $i$=$S$ to the
energetically favoured local structures.) Thus, active bond
concentration $S(\rb)$ is taken to have a frustrating influence on
crystallization at the local level, and each molecule
intrinsically has the cause of disorder and random fields against
the density ordering.

\subsection{Bond susceptibility}
\label{sec:2.2} Bond susceptibility is defined as a response
function measure of tendency for bond ordering or correlated
relaxation of bonds into their low-lying energy states, brought
about by the rearrangement of a molecular group/CRR. Bond
susceptibility apart from normalization is defined by \be \chi_b =
\left( \frac{\partial\langle M_b \rangle}{\partial H_b}
\right)_{T,H_b = 0} \label{eq:4.9} \ee where, $\langle M_b
\rangle$ denotes the thermal-averaged bond energy order parameter
characterizing the configurational energy of the system (more of
which in Sec.\ \ref{sec:3}), and the average field $H_b$ that is
referred to as the bond ordering field is a {\em self-generated}
molecular field favouring the local ordering of bonds and against
the density ordering or crystallization \cite{D01}. The above
physical quantities are introduced in order to be consistent with
the above two-order-parameter description of the liquids that
recognizes two competing ordering processes in a liquid, namely,
global density ordering that results in crystallization, and local
bond ordering that is responsible for glass transition. Bond
susceptibility is further expressed in terms of the equilibrium
fluctuations of the bond energy order parameter \be \chi_b =N\beta
\left\langle \delta m_b^2\right\rangle\,, \label{eq:5.3} \ee where
$\langle m_b \rangle = \langle M_b \rangle / N$ is the normalized
bond energy order parameter characterizing the configurational
energy, $\delta m_b = (m_b - \langle m_b\rangle)$ is the
corresponding fluctuation, $N$ is the system size, and angular
braces denote the usual thermal average. Eqs.\ (\ref{eq:4.9}) and
(\ref{eq:5.3}) for bond susceptibility can be readily derived from
the thermodynamic relation $ dG = -S\,dT - \langle M_b
\rangle\,dH_b$, which gives the change in free energy $G(T, H_b )$
of a system undergoing bond ordering as opposed to density
ordering or crystallization \cite{D01}. Evidently, the bond
ordering field $H_b$ is the thermodynamic conjugate-field that
couples to the bond energy order parameter $\langle M_b \rangle$
(which characterizes the configurational energy), and can be
regarded as the average concentration of energetically favoured
local structures $\bar{S}$.

\section{RELATIONS AMONG THERMODYNAMIC AND CORRELATION FUNCTIONS}
\label{sec:3} In this section the very concept underlying bond
susceptibility, i.e., bond ordering or correlated relaxation of
bonds, is generalized and applied to the case of laboratory
liquids where we treat intermolecular bonds as distinct objects
possessing internal degrees of freedom or energy states. The line
of reasoning presented culminates in identification of positional
specific heat as the thermodynamic response function associated
with the static relaxation length.

\subsection{Bond susceptibility and positional specific
heat} \label{sec:3.1} The bond energy order parameter $\langle m_b
\rangle$ is defined as a measure of the bond-order prevailing in a
system, and characterizes the configurational energy \cite{D01}.
By definition, it assumes large values when intermolecular bonds
are in their low-lying energy states as for a bond ordered low
temperature phase such as the glass, and is negligible when bonds
are distributed uniformly among all possible energy states that is
indeed the case when the thermal energy is far in excess of the
typical intermolecular binding energy. As a result, the bond
energy order parameter of a liquid in canonical (NVT) ensemble can
be simply defined in terms of configurational energy of the
liquid. With $\Phi (\rb_1 ,\rb_2 , \ldots , \rb_N )$ denoting the
potential energy function of a liquid composed of $N$ molecules,
the bond energy order parameter for this system is defined by \be
\langle m_b \rangle = - \langle\Phi\rangle /N. \label{eq:5.7} \ee
Eq.\ (\ref{eq:5.7}) satisfies all that is required of bond energy
order parameter. On substituting this expression into the
fluctuation-dissipation equation (\ref{eq:5.3}), for the liquid in
question we have \be \chi_b  = \beta \left\langle
\delta\Phi^2\right\rangle / N \label{eq:5.8} \ee where,
$\delta\Phi = (\Phi - \langle\Phi\rangle)$. Another response
function of interest and of immense relevance to the problem of
the glass transition is the specific heat, where for a liquid in
canonical ensemble may be expressed as a sum of two terms, a
kinetic part $C_k$, and an interaction or positional part $C_i$.
The above distinction follows from the fact that the liquid
Hamiltonian consists of two distinct parts: a kinetic energy part
$\sum_{i=1}^N \pb_i^2 / 2m_i$ covering the degrees of freedom
associated with the molecular momenta, and a potential energy part
$\Phi(\rb_1 , \rb_2 , \ldots , \rb_N)$ containing the
contributions to internal energy from interactions or positional
degrees of freedom. The positional part of the specific heat is
indeed the temperature rate of change of configurational energy:
\be C_i = \frac{1}{N} \frac{\partial\langle\Phi\rangle}{\partial
T}. \label{eq:5.05}\ee It can be readily shown that an expression
for the positional specific heat in terms of the equilibrium
fluctuations of the configurational energy is given by \be C_i /
k_B = \beta^2 \left\langle \delta\Phi^2 \right\rangle / N.
\label{eq:5.14} \ee On comparing Eq.\ (\ref{eq:5.8}) for the bond
susceptibility of a liquid in canonical ensemble with Eq.\
(\ref{eq:5.14}) for the positional part of the specific heat, we
arrive at the following simple result: \be \chi_b = T \cdot C_i\;.
\label{eq:5.15} \ee It  becomes evident from Eq.\ (\ref{eq:5.15})
that the bond susceptibility of a liquid in canonical ensemble,
characterizing the tendency for bond ordering or correlated
relaxation of bonds, can be simply interpreted as the response
function {\em positional specific heat}.

It is noteworthy that the result expressed by Eq.\ (\ref{eq:5.15})
is readily verifiable for certain lattice models such as the
two-dimensional Ising model \cite{KO}, and some impurity variants
thereof \cite{Thorpe}, where analytic solutions are available. In
particular, the four-spin correlation functions $w(r)=
\langle\sigma_1 \sigma_2 \sigma_r \sigma_{r+1}\rangle -
\langle\sigma_1 \sigma_2\rangle \langle\sigma_r
\sigma_{r+1}\rangle $ that can be also interpreted as {\em
two-bond energy correlation functions}, with $\sigma_1 \sigma_2$
characterizing the energy of a reference bond, while $\sigma_r
\sigma_{r+1}$ that of a bond in a different location in the
system, when summed over $r$ or all distinct pairs of bonds
essentially produce the specific heat  that is entirely
interaction or positional for the aforesaid lattice models:
$\sum_r w(r) =
\partial\epsilon /\partial\beta -1+\epsilon^2$, where $\epsilon =
\langle\sigma_1\sigma_2\rangle$ \cite{Thorpe}. This apparent
connection between bond susceptibility and two-bond energy
correlation functions will be used extensively next to establish a
quantitative relationship between bond susceptibility and static
relaxation length.

\subsection{Bond susceptibility and static relaxation length}
\label{sec:3.2} In the context of the bond ordering picture, a CRR
can be viewed as a correlated region of relaxing bonds. Thus, the
correlation length of such a region of bonds can be regarded as
the characteristic length of (cooperative) relaxation $\xi$.
Following Eq.\ (\ref{eq:5.3}), the bond susceptibility of a liquid
in NVT ensemble is expressed as \be \chi_b  = \frac{\beta}{V}
\left[ \langle M_b^2 \rangle - \langle M_b \rangle^2 \right]
\label{eq:5.16} \ee where, $V$ is the liquid volume, and $M_b$ is
the extensive bond energy parameter the thermal average of which
is the bond energy order-parameter characterizing the
configurational energy. For short-range molecular interactions,
which is almost always the case, $M_b$ can be expressed in terms
of the volume integral of a microscopic bond energy density
$m_b(\rb )=\sum_{i=1}^{N_b} m_{bi} \delta(\rb -\rb_i)$, where
$m_{bi}$ characterizes the energy of the $i$th bond, and $N_b$
denotes the total number of bonds in the system. Thus, for a
$d$-dimensional system we can write \be M_b = \int d^d r\;m_b(\rb
)\label{eq:5.17} \ee where, the integral is evaluated over the
liquid volume, and $m_b(\rb)$ characterizes the energy of an
intermolecular bond situated at $\rb$. On substituting this
expression into Eq.\ (\ref{eq:5.16}) and simplifying, we have
 \be \chi_b = \beta \int d^d
r\;\left[ \langle m_b (\rb )\;m_b (\Ob) \rangle\; - \;\langle m_b
(\rb )\rangle\;\langle m_b (\Ob)\rangle \right]. \label{eq:5.19}
\ee In Eq.\ (\ref{eq:5.19}), $m_b (\Ob)$ characterizes the energy
of a reference bond and $m_b (\rb )$ that of a bond at a distance
$r$ from the reference one. The quantity in the square brackets of
Eq.\ (\ref{eq:5.19}), is the two-bond energy correlation function
$G_b (\rb)\equiv\left\langle\;\delta m_b (\rb)\,\delta m_b
(\Ob)\;\right\rangle$ quantifying the spatial correlation of the
fluctuations of the bond energy order parameter. For an isotropic
system $G_b (\rb)=G_b (r)$. Furthermore, if we take the spatial
variation of $G_b(r)$ to be of the form \cite{BS,Fisher} \be G_b
(r) \sim \frac{g(r/\xi)}{r^{d-2}}, \label{eq:5.21} \ee where $\xi$
is a characteristic length beyond which the correlation function
rapidly vanishes; the bond susceptibility of a liquid and the
characteristic length of relaxation are thus related by \be
\chi_b\sim\beta\int^{\xi}dr\frac{r^{d-1}}{r^{d-2}}=\beta\;\xi^{2}/2.
\label{eq:5.22} \ee We must point out that in a previous work the
functional form of Eq.\ (\ref{eq:5.21}) has been used for a
similar spatial correlation function in the context of a defect
theory of relaxation to successfully recover the generalized
Vogel-Fulcher equation for the viscosity \cite{BS}. There the
relaxation is considered to be brought about by the movements of
mobile defects whose spatial correlation is governed by Eq.\
(\ref{eq:5.21}). In the context of the bond ordering picture, Eq.\
(\ref{eq:5.21}) is applied to the two-bond energy correlation
function $G_b (r)$ as the structural relaxation is now considered
to be a consequence of correlated relaxation of bonds within a
region whose average linear size gives the static relaxation
length $\xi$.

As a corollary of the result expressed by Eq.\ (\ref{eq:5.22}), we
note that a possible diverging bond susceptibility $\chi_b$ at
some finite temperature $T^*$ must necessarily imply a diverging
static relaxation length $\xi$ at the same temperature. That is,
if the variation with temperature of the bond susceptibility for a
system of interest is found to be a power law of the form $\chi_b
\sim (T - T^* )^{-\gamma_b}$, then the temperature variation of
the relaxation length must be also governed by a similar power law
$\xi \sim (T - T^*)^{-\nu}$ such that the exponents are related by
the scaling relation  \be \gamma_b = 2\nu\;.\label{eq:5.23}\ee We
must emphasize that the above result is consistent with a standard
result of statistical mechanics, namely, $\alpha = (2-\eta')\nu$,
where $\alpha$ is the specific heat exponent, and $\eta'$ is the
power law decay exponent of the energy-energy correlation function
\cite{Fisher}. In our  treatment leading to Eq.\ (\ref{eq:5.23}),
however, we have taken $\eta' = 0$ for a supercooled liquid system
that is corroborated by the numerical simulations of various
models of structural glass in three dimensions \cite{CCP,CC}. One
can further identify $\gamma_b$ with $\alpha$, as expected. As an
application to disordered systems, the results established here
will be used in Sec.\ \ref{sec:4} in a phenomenological
description for the static relaxation length of the fragile liquids.

\subsection{Positional specific heat and static relaxation
length} \label{sec:3.3} It has now become evident that the bond
susceptibility of a liquid can be expressed as $\chi_b = T \cdot
C_i$, where $C_i$ is that part of the specific heat containing
contributions from interactions or positional degrees of freedom.
Furthermore, bond susceptibility $\chi_b$ or indeed $C_i$ are
shown to be intimately related to the static relaxation length
$\xi$ such that essentially $\chi_b\sim C_i\sim\xi^2$ . With their
thermal behaviours so closely correlated, we therefore propose the
identification of positional specific heat $C_i$ as the
thermodynamic response function associated with the static
relaxation length $\xi$. Hence, we must further emphasize the
significance of the role played by the specific heat, and, in
particular, the interaction or positional part of it, in the
problem of the glass transition. Unfortunately not enough is known
about the precise behavior of the specific heat near $T_0$ and
much less about the interaction part of it, from an experimental
point of view, as the supercooled liquid falls out of equilibrium
on experimental time scales at kinetic glass temperature $T_g$
for the reasons pointed out in the Introduction. It is generally
believed that the excess specific heat over crystal value $\Delta
C_p$ that is regarded to be due to a subset of positional degrees
of freedom involving transitions between inherent structures (or
metabasins) of the potential energy hypersurface, rises with the
decreasing temperature in the supercooled temperature regime, and
a hyperbolic form $\Delta C_p \propto 1/T$ has been assumed in
conjunction with the Adam-Gibbs equation for the viscosity to
recover the Vogel-Fulcher equation \cite{Ang76}. However, a
drastically different $\Delta C_p$ has been also used to
accurately account for the viscosity of silicate glasses
\cite{Sch}. Hence, in the forthcoming section where the preceding
results will be applied to the case of fragile systems, a
semi-empirical approach is adopted to estimate the positional
specific heat of fragile supercooled liquids.

\section{Application to supercooled liquids}
\label{sec:4} In this section we present a phenomenological
description for the thermal dependence of the static relaxation
length for the general class of fragile liquids, in an attempt to
clarify some of the discrepancies that were referred to in the
Introduction.  As it turns out, the result obtained via this
phenomenological approach to a certain degree agrees with that
obtained from mean field theory of random-first-order transition,
also referred to in Sec.\ \ref{sec:1}, that suggests a power-law
temperature variation for the static relaxation length in the
quasi-equilibrium supercooled temperature regime with an apparent
divergence at the Vogel-Fulcher temperature $T_0$ \cite{Kirk}.

The positional specific heat $C_i$ is the temperature rate of
change of the configurational energy (Eq.\ (\ref{eq:5.05})). For
the case of fragile supercooled liquids where the $E_{\rm eff}$ of
Eq.\ (\ref{eq:5.25}) is to be largely interpreted as a
temperature-dependent `potential energy barrier' against the
viscous flow \cite{DH03,AG}, one can consider a relationship of
the form \be \langle\phi\rangle(T) \sim -E_{\rm eff}(T)
\label{eq:5.28} \ee where, $ \langle\phi\rangle =
\langle\Phi\rangle / N$ is the normalized configurational energy.
Eq.\ (\ref{eq:5.28}) expresses the average depth or minima of the
potential energy hypersurface explored by the liquid at each
temperature in terms of the height of the effective potential
energy barrier against the viscous flow. It should be also
mentioned that in the present analysis it is only the temperature
rate of change of the above quantities that is of interest. The
above equation simply indicates that the higher the energy barrier
$E_{\rm eff}$, the lower are the minima and consequently the
configurational energy $\langle\phi\rangle$. Eq.\ (\ref{eq:5.28})
is also consistent with a potential-energy-landscape
representation of a supercooled liquid according to which a liquid
is progressively confined to the deeper minima of the potential
energy hypersurface with the decreasing temperature whereby it
becomes more viscous due to an increased potential energy barrier
$E_{\rm eff}$ against the viscous flow \cite{DH03}, or
alternatively, a reduction in the configurational entropy
\cite{Goldstein}. Although this interpretation of the dynamics of
supercooled liquids in terms of increasing barrier heights with the 
decreasing temperature is not
the only one found in the current literature \cite{BG}, 
it is the one we deem appropriate for the purposes of the
present discussion. Hence,
the interaction or positional part of the specific heat of the
fragile liquids can be approximated by \be C_i = -\frac{\partial
E_{\rm eff}}{\partial T}\;. \label{eq:5.29} \ee Using the
effective potential energy barrier implied by the Vogel-Fulcher
equation (\ref{eq:5.25}), i.e., $ E_{\rm eff} = Ak_B T/(T-T_0)$,
we obtain \be \chi_b = T\cdot C_i = \frac{A k_B T
T_0}{(T-T_0)^2}\;. \label{eq:5.32} \ee Clearly, Eq.\
(\ref{eq:5.32}) implies a power law temperature variation of the
form $\chi_b \sim C_i \sim (T-T_0)^{-2}$ for bond susceptibility
as well as positional specific heat of the fragile supercooled
liquids, with an exponent $\gamma_b = 2$. From the scaling
relation Eq.\ (\ref{eq:5.23}), the exponent $\nu$ governing the
thermal variation of the static relaxation length of the fragile
supercooled liquids is thus given by $\nu = \gamma_b /2 = 1$.
Hence, with the effective potential energy barrier embodied in the
standard form of the Vogel-Fulcher equation, we obtain
\be \xi \sim (T-T_0)^{-1} \label{eq:5.34} \ee where, $T_0$ is
the Vogel-Fulcher temperature, and exponent $\nu$ is equal to
unity.

One may repeat the same analysis this time using the generalized
form of the Vogel-Fulcher equation: \be \eta = \eta_0\;\exp\left(
\frac{B}{(T-T_0)^{\gamma}} \right), \label{eq:5.35} \ee with the
identification, $E_{\rm eff} = Bk_B T/(T-T_0)^{\gamma}$, where $B$
and $\gamma$ are constant parameters. The special case of $\gamma
= 1$ results in the standard form of the Vogel-Fulcher equation
being recovered, however, different values for parameter $\gamma$
can be also found in the literature \cite{BS}. The following is
the result obtained with this rather generalized form of $E_{\rm
eff}$ that includes $\gamma$ as an extra parameter: \be \chi_b =
T\cdot C_i = \frac{Bk_BT}{(T-T_0)^{\gamma + 1}}\; \left[\gamma\;T-
(T-T_0)\right]\;. \label{eq:5.37} \ee In the limit $T\rightarrow
T_0$, Eq.\ (\ref{eq:5.37}) gives \be \chi_b(T\rightarrow T_0) \sim
\frac{\gamma\;Bk_B T_0^2}{(T-T_0)^{\gamma +1}}\;. \label{eq:5.38}
\ee Eq.\ (\ref{eq:5.38}) implies, $\chi_b \sim C_i \sim
(T-T_0)^{-(\gamma + 1)}$, and from Eq.\ (\ref{eq:5.23}), the
characteristic length exponent is now given by $\nu = (1 +\gamma
)/2$. Evidently an accurate experimental measurement of the
parameter $\gamma$ appearing in the generalized Vogel-Fulcher
equation is essential for a precise determination of the exponent
$\nu$ through the phenomenological procedure presented here.

A discussion of the observed difference between phenomenological
($\nu$=1) and mean field value ($\nu$=2/3) of the static
relaxation length exponent of the fragile liquids appears to be in
order at this stage. This difference can be attributed to the mean
field nature of the theory of random-first-order transition that
becomes exact in infinite dimensions, and is believed to have an
associated upper critical dimension $d_u=6$, which is
significantly higher than $d=3$ space dimensions of supercooled
systems. One therefore expects that the mean field estimate of the
relaxation length exponent becomes increasingly accurate as the
number of space dimensions approaches the upper critical value of
six. Thus, it appears that the mean field theory of
random-first-order transition, which presents qualitative features
analogous with the structural glass phenomenology, provides a
lower bound estimate of the static relaxation length exponent of a
supercooled liquid. Indeed, there are other instances where mean
field theories return lower estimates of correlation length
exponents. A prominent example is the mean field theory of the
continuous phase transitions (including the Ising model) that by
hyperscaling has an associated upper critical dimension of four,
which returns one-half for the correlation length exponent that is
again exceeded by the exact two-dimensional value (unity), and
reliable numerical estimates (0.63) for the corresponding
three-dimensional system. Nevertheless, It is a matter of
considerable interest that the phenomenological value of the
static relaxation length exponent $\nu = 1$, is precisely the
value obtained from the three-dimensional numerical simulations of
the microscopic models that exhibit random-first-order transition
in the mean field limit such as the $p$-spin glasses \cite{CCP},
and the frustrated Ising lattice gas model \cite{CC}.

In this section, a simple model for configurational energy in
terms of increasing barrier heights with the decreasing
temperature has been used that, despite simplicity, is applicable
to various types of fragile liquids with predominantly ionic, Van
der Waals, hydrogen, or covalent bonding. It would be also
interesting to look at certain specific models such as the
Rosenfeld-Tarazona relation for the Lennard-Jones liquid
\cite{RT}, or indeed any model that can be used to
distinguish between positional and kinetic contributions to the
specific heat in the context of the present work \cite{DSS}. That
effort is deferred to another work to be presented in due course.

\section{SUMMARY}
\label{sec:5} For a liquid in canonical ensemble it is shown that
the bond susceptibility and interaction or positional
part of the specific heat
are related by $\chi_b = T\cdot C_i$. Furthermore, bond
susceptibility and static relaxation length are found to vary as
$\chi_b \sim \beta\xi^2$. These relationships essentially point at
the identification of positional specific heat as the
thermodynamic response function associated with characteristic
length of relaxation---a proposition that further emphasizes the
significance of the role played by the positional specific heat in
the problem of the glass transition, and is likely to find further
applications in the theory of disordered systems, as applied here
to the case of fragile supercooled liquids in a phenomenological
description for the thermal dependence of static relaxation
length in those systems.

Through the phenomenological approach, the temperature variation
of the characteristic length of relaxation for fragile supercooled
liquids is determined to be governed by a power law $\xi \sim
(T-T_0)^{-\nu}$ implying an apparent divergence at the
Vogel-Fulcher temperature $T_0$, which to a certain degree agrees
with the corresponding result obtained from mean field theory of
random-first-order transition in that the apparent divergence
temperature is $T_0$ in both these cases. However, the
phenomenological exponent $\nu$ is found to be unity that is
higher than the corresponding mean field estimate, hence, favouring
a stronger temperature dependence for the static relaxation length
in the supercooled temperature regime. This difference can be
attributed to the mean field nature of the theory of
random-first-order transition, as discussed above. It is indeed a
matter of considerable interest that the phenomenological exponent
$\nu = 1$ is in perfect agreement with the corresponding value
obtained from the three-dimensional numerical simulations of the
same models of structural glass that exhibit random-first-order
transition in the mean field limit.

\end{document}